\documentclass[lang=en]{elegantpaper}
\usepackage[misc]{ifsym}
\usepackage{hyperref}
\usepackage[shortlabels]{enumitem}
\usepackage[utf8]{inputenc}
\usepackage{booktabs}
\usepackage[justification=centering]{caption}
\usepackage{amsmath,graphicx}
\usepackage{pgfplots}
\usepackage{tikz}

\usetikzlibrary{positioning, shapes, fit, arrows}
\usetikzlibrary{external}

\title{Variational Auto-Encoder based Mandarin Speech Cloning}
\author{Xing Qingyu\textsuperscript{\footnote{Xing proposed using VAENAR-TTS in CV2TTS, worked on implementing the idea, and experimented with novel subjective tests.} \ $^{\textrm{\Letter}}$}\\Yanzhou No.1 Senior High School\\\email{xqy@tts.ac.cn}		\and
        Ma Xiaohan\textsuperscript{\footnote{Ma organized participation in new dataset and as a consultant for the broadcast program.}}\\Yanzhou No.1 Senior High School\\\email{mxh@tts.ac.cn}}

\begin{document}

\maketitle
\begin{abstract}
Speech cloning technology is becoming more sophisticated thanks to the advances in machine learning. 
Researchers have successfully implemented natural-sounding English speech synthesis and good English speech cloning by some effective models. 
However, because of prosodic phrasing and large character set of Mandarin, Chinese utilization of these models is not yet complete. 
By creating a new dataset and replacing Tacotron synthesizer with VAENAR-TTS, we improved the existing speech cloning technique CV2TTS to almost real-time speech cloning while guaranteeing synthesis quality. 
In the process, we customized the subjective tests of synthesis quality assessment by attaching various scenarios, so that subjects focus on the differences between voice and our improvements maybe were more advantageous to practical applications. 
The results of the A/B test, real-time factor (RTF) and 2.74 mean opinion score (MOS) in terms of naturalness and similarity, reflect the real-time high-quality Mandarin speech cloning we achieved.
\end{abstract}

\section{Introduction}
Many models have been developed for text-to-speech (TTS) such as WaveNet vocoder \cite{oord2016wavenet} and Tacotron synthesizer \cite{wang2017tacotron}, which helps to move speech synthesis away from mechanized timbres. 
And integrated end-to-end TTS systems which only need correspondence labeled between text and audio rather than complex pipelines are already widely used \cite{wang2017tacotron}\cite{zen2007hmm}. 
Not only speech synthesis, but also speech cloning has made great progress in recent years, and it is even possible to synthesize new speech by a few seconds of untranscribed reference audio \cite{CV2TTS}. 
But we noted that speech cloning is not as widely used as computer vision related technology now, especially in Chinese, so it could be an important direction to be developed. 
For example, it can be used for people who have lost voice to communicate naturally and real-time voiceover work in games, greatly reducing the amount of storage space, quickly modifying text messages with less loss.\par
Based on CV2TTS \cite{CV2TTS}, we proposed CV2TTS-VAENAR to handle prosodic phrasing and large character set of Mandarin, trying to provide a method for Mandarin speech cloning. 
In the process that we used the subjective tests to reflect quality of speech synthesis and cloning included MOS \cite{VISWANATHAN200555} and A/B test \cite{hirst1998comparison}, and found that they generally didn't include factors in the application scenario, so the results thus obtained are indeed universal, but inappropriate in some application scenarios. 
For subjective tests, adding more scenario factors maybe is more conducive to the use of specialization techniques and more effectively reflect the adaptability of real scenarios.\par
Naturally, regulation was always going to lag behind technological development. 
Although speaker verification schemes for speech cloning already existed even it was the origin of CV2TTS \cite{CV2TTS}, and Subramani et al. \cite{subramani2020learning} also proposed efficient representations for the problems. But in the practical application of speech cloning there are currently no relevant industry specification. 

\section{Related work}
\begin{figure}[h]
    \centering
    \begin{tikzpicture}[auto, font=\small, node distance=0.7cm and 0.5cm, >=latex']
      \pgfdeclarelayer{back}
      \pgfsetlayers{back,main}
  
      \tikzstyle{block} = [draw, fill=blue!20, align=center, rectangle, minimum height=2em, minimum width=4.5em]
      \tikzstyle{speakerenc block} = [block, fill=green!20]
      \tikzstyle{vocoder block} = [block, fill=red!20]
      \tikzstyle{input} = [align=center, inner sep=0]
      \tikzstyle{output} = []
  
      \node [input, name=inputaudio, align=center, font=\scriptsize\sffamily] {speaker \\ reference \\ waveform};
      \node [speakerenc block, right=0.8cm of inputaudio, inner ysep=1ex] (speakerenc) {Speaker \\ Encoder};
      
      \node [input, name=inputtext, below= of inputaudio, align=center, font=\scriptsize\sffamily] {grapheme or \\ phoneme \\ sequence};
      \node [block] (ttsenc) at (inputtext -| speakerenc) {Encoder};
      \node [block, right=0.25cm of ttsenc, minimum width=0] (concat) {concat};
      \node [block, right=0.25cm of concat] (ttsatt) {Attention};
      \node [block, right=0.25cm of ttsatt] (ttsdec) {Decoder};
      \node [above=0.09cm of ttsenc, inner sep=0, align=right, font=\small] (synthlabel) {\hspace{-0.5ex}Synthesizer};
      \begin{pgfonlayer}{back}
        \node [fill=blue!10, fit=(synthlabel)(ttsenc)(ttsdec), inner xsep=0.08cm] {};
      \end{pgfonlayer}
      \node [vocoder block, right=0.7cm of ttsdec] (vocoder) {Vocoder};
      \node [output, right= of vocoder, font=\scriptsize\sffamily] (output) {waveform};
  
      \draw [->] (inputaudio) -- (speakerenc);
      \draw [->] (speakerenc) -| node[align=left, font=\scriptsize\sffamily, color=darkgray] {speaker \\ embedding} (concat);
      
      \draw [->] (inputtext) -- (ttsenc);
      \draw [->] (ttsenc) -- (concat);
      \draw [->] (concat) -- (ttsatt);
      \draw [->] (ttsatt) -- (ttsdec);
      \draw [->] (ttsdec) edge[bend left=45] (ttsatt);
      \draw [->] (ttsdec) -- node [align=left, font=\scriptsize\sffamily, pos=1.2, color=darkgray] {log-mel \\ spectrogram \\ \vspace{1ex}} (vocoder);
      \draw [->] (vocoder) -- (output);
    \end{tikzpicture}
    \caption{CV2TTS Model overview.}
    \label{fig.model_overview}
  \end{figure}
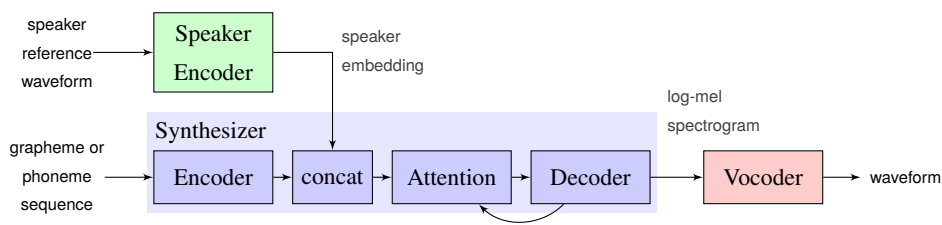

There were some existing works that applied speech synthesis of various languages \cite{krstulovic2007hmm}\cite{karabetsos2008hmm} which was based on existing models such as Hidden Markov Models (HMM) and Tacotron, and part with our work is similar to them, which appears to be closer to engineering. 
These works were attempts to generalize the model, and as Chinese is the largest native language in the world \cite{spokenlanguage}, this generalization has a more important meaning. 
On the basis of speech synthesis, we learned about the idea of speech cloning \cite{CV2TTS}, and the basic speech cloning model CV2TTS is illustrated in Figure~\ref{fig.model_overview}.
\par
The system consists of three independently trained neural networks, which is also the dominant speech cloning scheme:
\begin{enumerate}
\item It proposed a scalable and accurate neural network framework for speaker verification, in which a training task for speaker discrimination yields an embedding that can be used to condition the synthesis network on speaker identity.
\item It extended the recurrent sequence-to-sequence with attention Tacotron 2 \cite{shen2018natural} architecture to support multiple speakers following a scheme similar to the method used by Gibiansky et al. \cite{gibiansky2017deep}. 
\item It used the sample-by-sample autoregressive WaveNet \cite{oord2016wavenet} as a vocoder to invert synthesized mel-spectrograms emitted by the synthesis network into time-domain waveforms.
\end{enumerate}
\hspace*{\fill}\par
Many papers have been published in which some of the models used in this system have been improved. 
Weiss et al. \cite{9413851} proposed a Tacotron-based synthesizer model which uses a fixed intermediate representation, and learns all parameters end-to-end. 
Its experiments showed that the synthesizer model generates speech with quality approaching a state-of-the-art neural TTS system, with improved generation speed.\par

In addition to its work on the Tacotron model, FastSpeech \& FastSpeech 2/2s are popular synthesizer models because of its non-autoregressive feature. 
FastSpeech \cite{NEURIPS2019_f63f65b5}, which represents non-autoregressive TTS in comparison to autoregressive Transformer TTS, can synthesize speech significantly faster and with comparable quality than previous autoregressive models. 
FastSpeech 2 \cite{ren2021fastspeech} is the next generation of FastSpeech, which trains the model directly with ground-truth targets rather than the simplified output from the teacher, and introduces more speech variation information (e.g., pitch, energy, and more accurate duration) as conditional inputs, resulting in better results. 
Indeed, similar work has been published using FastSpeech 2 for Mandarin speech synthesis and cloning, with good results \cite{9497934}.\par

However, to generate a hard alignment between the text and the spectrogram, FastSpeech (2) models rely on phoneme-level durations named (Auto-Regressive) AR teacher model which is time-consuming to obtain duration labels. 
Worse, hard alignment based on phoneme expansion has the potential to degrade the naturalness of synthesized speech. 
VAE (Variational Auto-Encoder) has been widely used in voice conversion \cite{9367139} and speech synthesis \cite{Cong2021GlowWaveGANLS} because of its effectiveness.
Our speech cloning system employs the VAENAR-TTS synthesizer model \cite{lu2021vaenar} to achieve good result.

\section{Architecture}
\begin{figure}[h]
  \centering
  \includegraphics[width=0.5\linewidth]{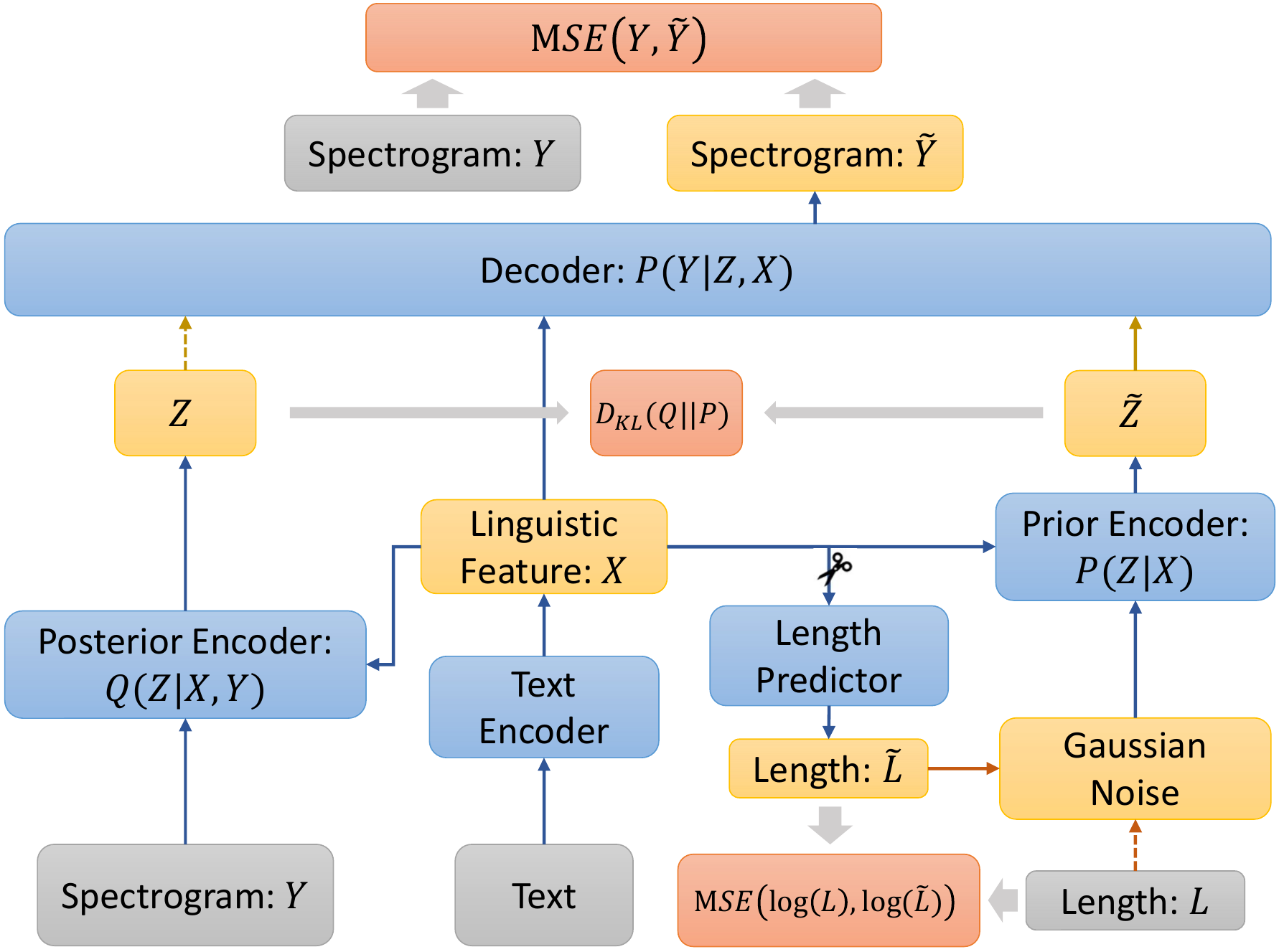}
  \caption{Architecture of VAENAR-TTS. The dotted lines are only turned on during the training stage, while their counterparts with the same color are only turned on during the inference phase. The scissors denote that the gradients for the length predictor will not be back-propagated to the text encoder. }
  \label{fig:synthesizermodel_architecture}
\end{figure}

To obtain real-time voice cloning, we used the CV2TTS-like architecture whose each of the three components are trained independently:
\begin{enumerate}
\item A recurrent speaker encoder to compute a fixed dimensional vector from a speech \cite{Wan2018GeneralizedEL}.
\item A extended non-autoregressive sequence-to-sequence synthesizer VAENAR-TTS to predict a mel-spectrogram by speaker embedding vector. Its architecture is shown in Figure~\ref{fig:synthesizermodel_architecture} which can be readily derived with \ref{equ:vaeformal} formalization \cite{lu2021vaenar}.
\item An non-autoregressive vocoder HiFi-GAN to convert the mel-spectrogram into time domain waveforms \cite{kong2020hifigan}.
\end{enumerate}
\begin{align}
  \log[Pr(Y|X)]\approx &-\int_{Z}{Q(Z|X,Y)\log[\frac{Q(Z|X,Y)}{P(Z|X)}]dZ} \nonumber + \int_{Z}{Q(Z|X,Y)\log[P(Y|Z,X)]dZ}\\
  \label{equ:vaeformal}
\end{align}
\par
VAENAR-TTS synthesizer plays a great role in the architecture. 
Specifically, unlike Glow-based NAR-TTS models \cite{kim2020glow} that used it as the decoder to predict the spectrogram, the model used Glow \cite{kingma2018glow} to model the prior distribution of the latent variable.
\subparagraph{Loss Function}
We use the standard loss function.\par
The vocoder training loss function of HiFi-GAN is shown below:
\begin{align}
  \mathcal{L}_{G} &= \sum_{k=1}^{K}\Bigg[\mathcal{L}_{Adv}(G; D_k) + \lambda_{fm}\mathcal{L}_{FM}(G; D_k)\Bigg] + \lambda_{mel}\mathcal{L}_{Mel}(G)\\
  \mathcal{L}_{D} &= \sum_{k=1}^{K}\mathcal{L}_{Adv}(D_k; G)
  \label{equ:hifigantrainingloss}
\end{align}\par
The synthesizer training loss function of VAENAR-TTS is shown below:
\begin{align}
  \mathcal{L} = &\mathbf{MSE}(Y, \tilde{Y}) + \alpha \mathbf{D_{KL}}(Q(Z|X,Y)||P(Z|X)) \nonumber\\
  &+ \beta \mathbf{MSE}(\log(L), \log(\tilde{L}))
\label{equ:synthesizertrainingloss}
\end{align}

\section{Experiments}
The Experiments part includes training and subjective tests.

\subsection{Training}
\begin{figure}[h]
  \centering
  \includegraphics[width=0.5\linewidth]{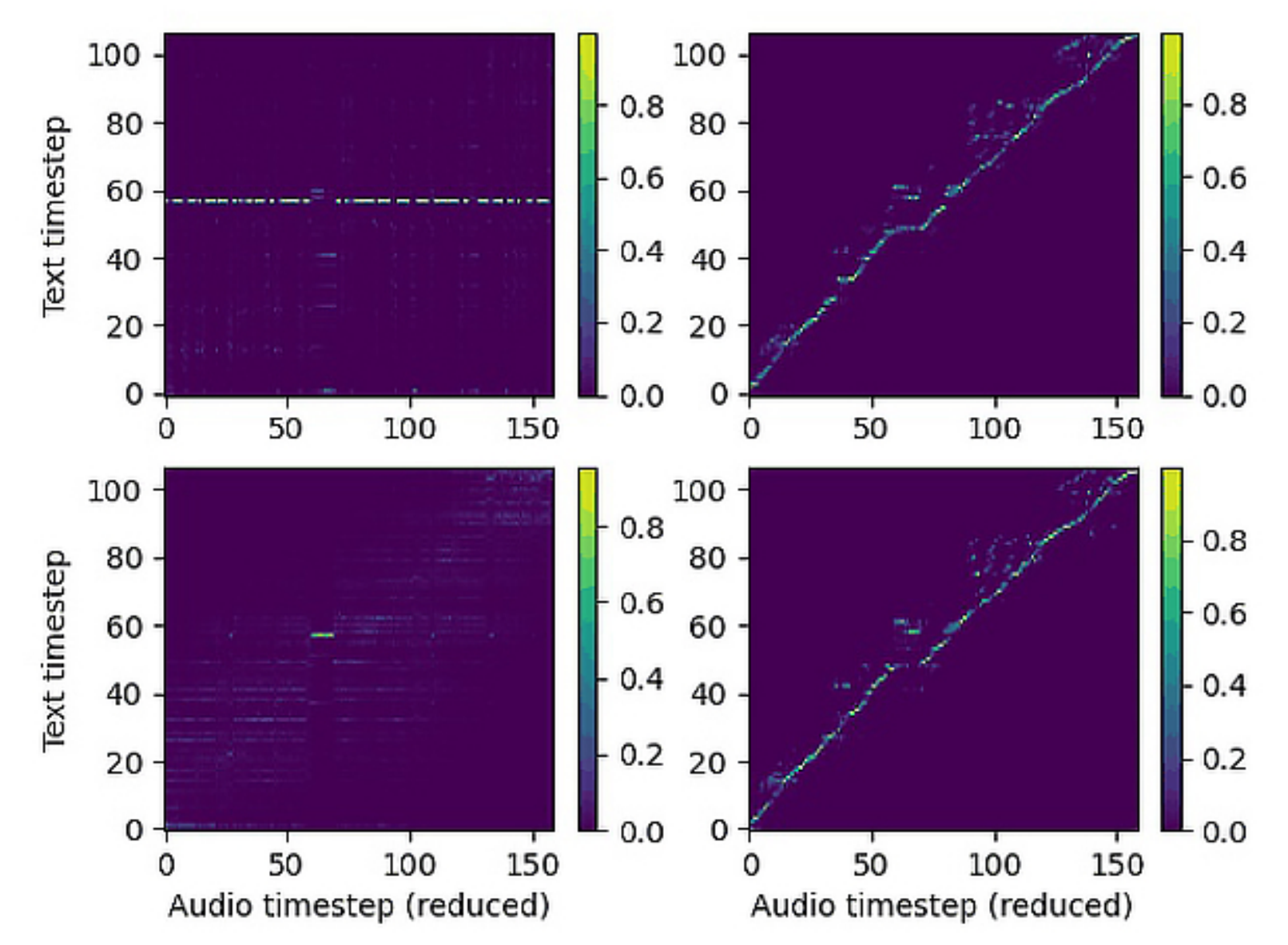}
  \caption{Two examples of audio timestep at 70K steps.}
  \label{fig:audiotimestep}
\end{figure}

We refer to the code implementation of a voice cloning project named MockingBird \cite{mocking-bird} and PyTorch implementation of VAENAR-TTS \cite{lee2021vaenar-tts}. 
We follow the official VAENAR-TTS training parameters nearly to train synthesizer model on a NVIDIA Geforce TITAN Xp GPU. 
In this process, the VAENAR-TTS model has excellent alignment, illustrated in Figure~\ref{fig:audiotimestep}.

\subparagraph{Datasets}
The official VAENAR-TTS implementation only used the Databaker Mandarin dataset, which only has female voices. 
To obtain better synthesis and cloning quality, we train the system on larger datasets, including aidatatang\_200zh, magicdata, aishell1 \cite{aishell_2017} and aishell3 \cite{shi2021aishell} datasets from OpenSLR and our private dataset. 
The mixed datasets total over 1200 hours.\par
Our private dataset was created using 48 kHz, 16-bit audio data recorded from Sennheiser MK 4 high-fidelity microphone and re-sampled to 16 kHz, 16-bit WAV format, which is the mainstream technology. 
The entire audio recording process was filmed by Xiaomi 10 for verification and remediation. 
In quality checking stage which is general steps of dataset creation:
\begin{enumerate}
\item Economic and time constraints, different from aishell3 \cite{shi2021aishell}, we first use Automatic Speech Recognition (ASR) to initially mark the recorded audio, then check and remove utterances with inconsistent raw text and transcription manually.
\item Text normalization (TN) is carefully applied towards English words, numbers, names, places, streets, organizations, abbreviations, brands and shops, for example:
\begin{itemize}
\item Number: 123 are normalized to yi1 er4 san1.
\item URL: the ``www.abc.com'', are normalized to ``san1 W dian3 A B C dian3 com''.
\item English abbreviations: VIP, CCTV are presented in uppercase.
\end{itemize}
\item utterances containing obvious mis-pronunciations are removed.
\item Some additional tagging for future work (e.g., scenarios, emotions and overviews).
\end{enumerate}\par
Private dataset was produced with the consent of the speaker, but unfortunately we cannot make it public in addition to the samples used for experiments due to the agreement.

\subsection{Subjective Tests}
The original subjective tests claimed that subjects were unaffected by other factors, an ideal situation that is difficult for our team to achieve and difficult to verify that it was achieved \cite{hirst1998comparison}. 
Therefore, we propose a novel subjective test. \par
The MOS scores are listed in Table~\ref{tab:mos}. 
The results show that the MOS scores of proposed CV2TTS-VAENAR higher than the other models with excellent RTF. 
The difference in MOS scores is also confirmed in the A/B test between CV2TTS-VAENAR and CV2TTS in Table~\ref{tab:ab_test}.

\begin{table}[htb]
  \caption{The MOS with 95\% confidence intervals for the improved system named CV2TTS-VAENAR, ground-truth samples (GT), CV2TTS-Glow and CV2TTS baseline.}
  \centering
  \begin{tabular}{c c c}
  \toprule
       Model & MOS(naturalness \& similarity) & RTF\\
       \midrule
       GT & 4.66 \(\pm\) 0.11 & -\\
       \midrule
       CV2TTS & 2.43 \(\pm\) 0.11 & $1.78\times10^{-1}$\\
       CV2TTS-Glow & 2.53 \(\pm\) 0.12 & $1.26\times10^{-2}$\\
       \textbf{CV2TTS-VAENAR} & \textbf{2.74} \(\pm\) 0.09 & \textbf{$9.85\times10^{-3}$}\\
       \bottomrule
  \end{tabular}
  \label{tab:mos}
\end{table}
\begin{table}[htb]
\caption{Results of preference test between CV2TTS-VAENAR and CV2TTS on speaker. The evaluation is at \textit{p} \textless 0.01 level.}
    \centering
    \begin{tabular}{c c c c}
    \toprule
         & \textbf{CV2TTS-VAENAR} & CV2TTS & Same \\
         \midrule
         Preference \% & \textbf{64.375} & 15.625 & 20.000 \\
         \bottomrule
    \end{tabular}
    \label{tab:ab_test}
\end{table}
\subparagraph{Experimental Setup}
We adapt the traditional subjective test environment to combine it with practical applications to better reflect quality of speech synthesis and cloning \cite{jones1999organizational}. 
This subjective test contains twenty sentences which were not included in the training data with a variety of scenarios in Table~\ref{tab:sampleoverview}.
\begin{table}[htb]
  \caption{Scenarios, overviews and number of sample sentences}
      \centering
      \begin{tabular}{c c c}
      \toprule
           Scenarios & Overviews & Number \\
           \midrule
           Daily Conversations & Dating, Current Affairs and Memories  & 4 \\
           \midrule
           News Broadcast & Earthquake and Epidemic & 2 \\
           \midrule
           Public Broadcast & Lost \& Found and Train Terminus & 2 \\
           \midrule
           Human Customer Service & Welcome and Response & 2 \\
           \midrule
           Phrase Read & Book Quotes & 2 \\
           \midrule
           Game Voiceover & Genshin Impact Copywriting & 2 \\
           \midrule
           Guided Tour & Kong Mansion and Forbidden City & 2 \\
           \midrule
           Faculty Teaching & Marxism and Linear Algebra & 2 \\
           \midrule
           Whisper & Comforting & 2 \\
           \bottomrule
      \end{tabular}
      \label{tab:sampleoverview}
  \end{table}
\hspace*{\fill}\par
We evaluate our proposed approach on speech cloning by training three models on the Datasets paragraph described above. 
They all use the state of the art GAN-based vocoder HiFi-GAN whose implementation is public for a fair comparison.\par 
The synthetic speech's naturalness and similarity were assessed using the MOS test and the A/B test. 
The test included 16 native listeners, and the speech samples were shuffled in each test.\par
The real-time factor (RTF), which is the time it takes to synthesize one second of speech spectrogram from text, is used to determine synthesis speed. 
The RTF benchmarks are run on a single NVIDIA Geforce TITAN Xp GPU with a single batch size which averaged over 10 times runs on the entire test set for each model.

\section{Conclusions and Discussions}
We present a VAENAR-TTS-based Mandarin speech cloning system similar to CV2TTS, which demonstrates the feasibility of porting the alignment learning synthesizer model to speech cloning system. 
Testing results show that compared to Tacotron-based CV2TTS, CV2TTS-VAENAR has higher naturalness, similarity and real-time performance in Mandarin speech cloning as demonstrated Figure~\ref{tab:mos}~\ref{tab:ab_test}. 
Our innovations in subjective test which introduce scenarios of the test are theoretically valid to Multi-Scenario applications and can be further investigated.\par
Here we propose a idea for daily conversations benchmark test refered to the traditional The Turing Test \cite{oppy2003turing}. We are unable to implement this test now, but we guess it will work. 
It is related to The Total Turing Test proposed by Harnad \cite{harnad1989minds}\cite{harnad1991other} who considered that a better test than The Turing Test will be one that requires responses to all of our inputs. 
Construct a robot with something like human sensorimotor capabilities is not the scope of our study for this paper and we don't even care if participant is a computer (e.g., English GPT-3 \cite{floridi2020gpt} and Chinese CPM-2 \cite{zhang2021cpm})or a human. 
However, we can explore the correlation of the results of the test with different TTS methods to confirm the advantage of a certain method in practical applications. \par

We notice that PAMA-TTS \cite{he2021pama} is another model of alignment learning which applies monotonic constraint rather than leverages both annealing reduction factor and causality mask to help attention-based alignment learning so we are curious if it will work better in the field of speech cloning. 
Although we achieve better speech cloning quality, the expression of tone and is essential in the field of voiceover and in communication. And incorporating expressiveness into speech synthesis systems makes listeners more enjoyable to listen to and interact with has been proven \cite{henter2017principles}. 
Akuzawa et al. \cite{akuzawa2018expressive} proposed a method for combining VoiceLoop and VAE into the speech synthesis architecture resulting in higher quality speech, may be applied to the next step in the development of Mandarin synthesis.\par
Furthermore, with the development of AI accelerators for mobile devices, real-time speech cloning with autoregressive models may be possible, resulting in better results. 
Finally we demand that industry regulations and legislation on speech synthesis and cloning be improved and this aspect should be discussed by legal experts to promote the healthy development of the whole industry.

\section{Acknowledgements}
All of the work for this paper was done outside of school hours and even during Chinese New Year. 
We are both studying at Yanzhou No.1 Senior High School and would like to thank the school teachers for our support in basic education, especially Ma Xincun and Zhao Yongjun.\par
Even though all of the "innovations" are obvious in the paper, doing the majority of the work in the speech cloning process by ourselves, which included Computer Science, Sound, Engineering, Broadcasting, and even Psychology at a time when interdisciplinary research is becoming mainstream, is rare and meaningful. 
Every computer scientist has aspired to be cross-disciplinary experts like Alan Mathison Turing and John von Neumann, and we are no exception. 
If at all possible, we will work in that direction from now on.


\bibliography{cite}

\newpage

\begin{appendix}
\section{Subjective tests screenshots}
\begin{figure}[h]
  \centering
  \includegraphics[width=0.9\linewidth]{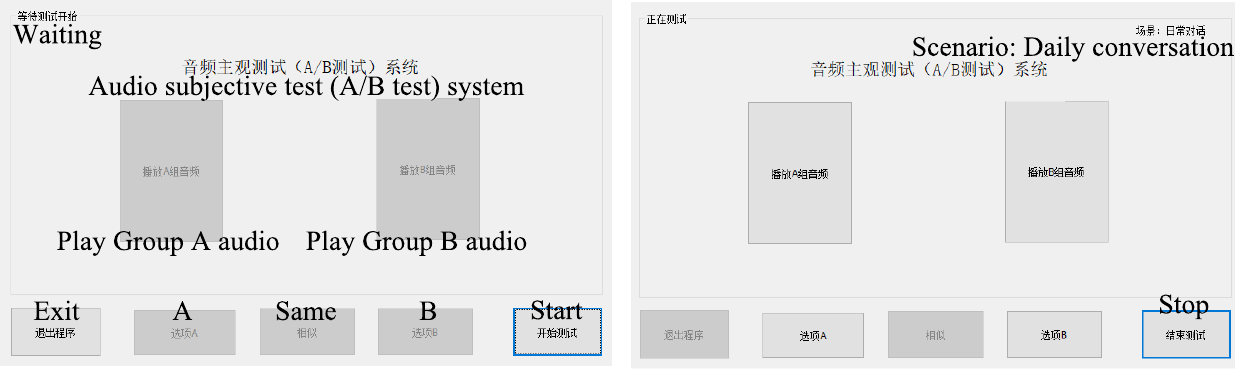}
  \caption{Graphical User Interface (GUI) of subjective test system, reflecting our design philosophy for scenarios.}
  \label{fig:ABTESTGUI}
\end{figure}
\section{Formula Definition}
\subsection{HiFi-GAN}
\begin{align}
  \mathcal{L}_{Adv}(D; G) &= \mathbb{E}_{(x, s)} \Bigg[(D(x)-1)^2 + (D(G(s)))^2\Bigg]\\
  \mathcal{L}_{Adv}(G; D) &= \mathbb{E}_{s} \Bigg[(D(G(s))-1)^2\Bigg]\\
  \mathcal{L}_{Mel}(G) &= \mathbb{E}_{(x, s)} \Bigg[||\phi(x)-\phi(G(s))||_{1}\Bigg]\\
  \mathcal{L}_{FM}(G; D) &= \mathbb{E}_{(x, s)} \Bigg[\sum_{i=1}^{T}\frac{1}{N_{i}}||D^i(x)-D^i(G(s))||_{1}\Bigg]\\
  \mathcal{L}_{G} &= \mathcal{L}_{Adv}(G; D) + \lambda_{fm}\mathcal{L}_{FM}(G; D) + \lambda_{mel}\mathcal{L}_{Mel}(G)\\
  \mathcal{L}_{D} &= \mathcal{L}_{Adv}(D; G)
\end{align}
\subsection{VAENAR-TTS}
\begin{align}
  P(Y|Z,X)=\prod_{i=1}^{N}{P(y_i|Z,X)}
\end{align}
\end{appendix}

\end{document}